\documentclass{egpubl}
\usepackage{eg2026}
 
\ConferenceSubmission   
\ShortPresentation      
\usepackage[T1]{fontenc}
\usepackage{dfadobe}  

\usepackage{cite}  
\BibtexOrBiblatex

\electronicVersion
\PrintedOrElectronic
\ifpdf \usepackage[pdftex]{graphicx} \pdfcompresslevel=9
\else \usepackage[dvips]{graphicx} \fi

\usepackage{egweblnk} 

\usepackage{amsmath}
\usepackage{amssymb}

\usepackage{tikz}
\usepackage{graphicx}
\usetikzlibrary{arrows.meta}

\usepackage{pgfplots}
\pgfplotsset{compat=1.18}

\title[Voxel Deformation-Aware Neural Intersection Function]%
      {Voxel Deformation-Aware Neural Intersection Function}

\author[C.C. Kao \& G. Makowski \& S. Fujieda \& T. Harada]
{\parbox{\textwidth}{\centering 
        Chih-Chen Kao \orcid{0000-0002-7631-2284}, 
        Grzegorz Makowski \orcid{0009-0008-4945-8246}, 
        Shin Fujieda \orcid{0000-0002-2472-7365},
        Takahiro Harada \orcid{0000-0001-5158-8455}
        }
\\
{\parbox{\textwidth}{\centering Advanced Micro Devices, Inc. (AMD)
       }
}
}

\begin{document}

\teaser{
 \includegraphics[width=0.95\linewidth]{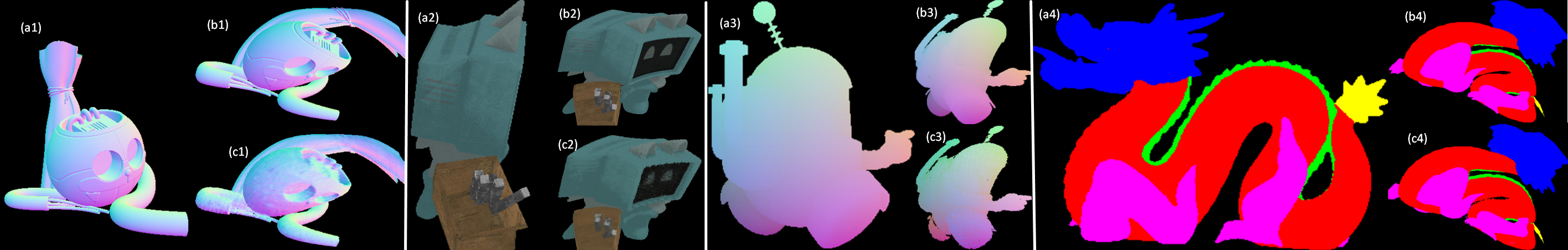}
 \centering
  \caption{Voxel Deformation-Aware Neural Intersection Function predicting geometric properties under parameterized deformation. From left to right: normal, albedo, distance, and material. For each example: (a) original shape, (b) deformed reference, (c) predicted result.}
\label{fig:teaser}
}

\maketitle


\begin{abstract}

We extend the Locally-Subdivided Neural Intersection Function (LSNIF) to support parameterized deformable and animated geometry. Our approach introduces a rest-space and deformed-space formulation inspired by meshless rendering, allowing ray samples to be mapped back to a canonical space where a single neural network represents geometry consistently across poses without retraining. To maintain accuracy under deformation-aware training, we incorporate scale-invariant distance regression, uncertainty-weighted multi-task learning, and a hybrid positional-grid encoding. The resulting method preserves the compactness and efficiency of LSNIF while enabling robust neural intersection prediction for dynamic geometry.


\end{abstract}

\section{Introduction}

Neural representations for geometry and intersection queries have recently demonstrated impressive efficiency and accuracy, offering a promising alternative to traditional acceleration structures~\cite{10.1145/3528223.3530127}. Recent work such as N-BVH~\cite{weier2024n} further explores hybrid approaches that combine neural ray queries with classical bounding volume hierarchies, highlighting a spectrum between learned and explicit acceleration structures. In particular, the Locally-Subdivided Neural Intersection Function (LSNIF) showed that object-space intersection queries can be represented by a compact neural network trained on voxelized geometry~\cite{10.1145/3728295}. By encoding spatial information using a sparse hash grid, LSNIF predicts intersection distance, surface normal, albedo, material, and occlusion for arbitrary query rays. This design enables accurate and memory-efficient intersection prediction while eliminating the need for a bottom-level Bounding Volume Hierarchy (BVH).

Despite these advantages, existing neural intersection methods, including LSNIF, are limited to static geometry. In practice, many scenes contain deformable or articulated objects whose shape or pose varies over time. Handling deformation efficiently remains challenging, as neural networks trained on static geometry do not generalize to unseen poses without retraining. Related work in neural rendering has addressed dynamic scenes using deformation-aware implicit representations, such as D-NeRF~\cite{pumarola2021d} and Deformable NeRF~\cite{park2021nerfies}, which learn a mapping between canonical and posed spaces. Extensions such as HyperNeRF~\cite{park2021hypernerf} further increase representational capacity for complex topological changes. However, these approaches focus on volumetric rendering rather than explicit ray-surface intersection queries, and are not directly designed for efficient object-space querying.

Classical graphics pipelines address deformation through skinning, meshless deformation, or volumetric warping by defining a mapping between a canonical rest space and a deformed space. However, incorporating such deformation models into neural implicit representations remains non-trivial, since geometry is encoded implicitly within network parameters.

To address this limitation, we extend the LSNIF framework to support parameterized deformation of voxelized geometry by adopting a rest-space and deformed-space formulation inspired by meshless rendering methods~\cite{luton:hal-05095359}. The neural network is trained exclusively in rest space, while ray tracing and data generation are performed in deformed space. Ray samples obtained in deformed space are mapped back to rest space before being encoded and processed by the network. This allows a single neural model to represent geometry consistently across multiple poses, enabling efficient and accurate intersection prediction for animated or deformable objects without per-pose retraining.

To maintain robustness when training across a wide range of deformations, we introduce deformation-aware training strategies that stabilize distance prediction, balance multiple geometric and appearance objectives, and improve spatial expressiveness under pose variation. In combination, these contributions yield a deformation-aware neural intersection function that preserves the compactness and efficiency of LSNIF while enabling robust and accurate intersection prediction for dynamic geometry.

\begin{figure*}[t]

\centering
\begin{tikzpicture}[>=Stealth, thick]

\begin{scope}[xshift=0\textwidth]

\node[anchor=south west, inner sep=0] (bunnyA) at (0,0)
  {\includegraphics[width=3cm]{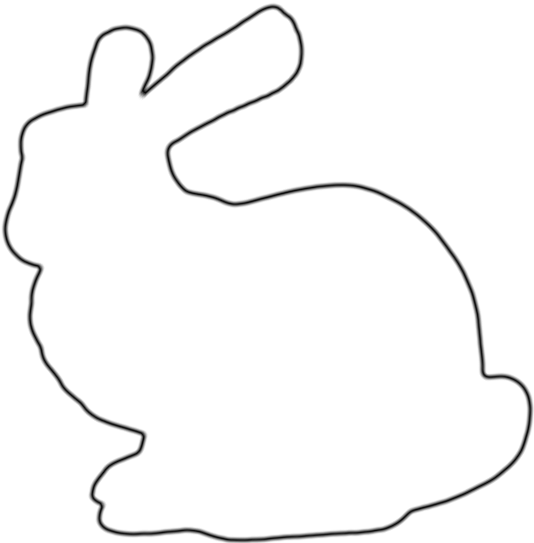}};

    \begin{scope}[x={(bunnyA.south east)}, y={(bunnyA.north west)}]
    \draw[orange, line width=0.6pt] (0,0) grid[step=0.1] (1,1);
    \end{scope}
\end{scope}

\draw[->, gray, line width=1.2pt] (3.5,1.5) -- (4.2,1.5);

\begin{scope}[xshift=0.25\textwidth]

\node[anchor=south west, inner sep=0, opacity=0] (bunnyB) at (0,0)
  {\includegraphics[width=3cm]{bunny_small.png}};

    \begin{scope}[x={(bunnyB.south east)}, y={(bunnyB.north west)}]
    
    \def\gridSize{10} 
    \def\gridHeight{10} 
    \def\maxRotation{5} 
    \def\originX{0.0} 
    \def\originY{0.5} 
    
    \foreach \x in {0,...,\gridSize} {
      \pgfmathsetmacro{\theta}{\x * \maxRotation / \gridSize}
      \foreach \y in {0,...,\gridHeight} {
        \pgfmathsetmacro{\origXnorm}{\x/\gridSize}
        \pgfmathsetmacro{\origYnorm}{\y/\gridHeight}
        
        \pgfmathparse{cos(\theta)*(\origXnorm-\originX) - sin(\theta)*(\origYnorm-\originY) + \originX}
        \let\rotX\pgfmathresult
        \pgfmathparse{sin(\theta)*(\origXnorm-\originX) + cos(\theta)*(\origYnorm-\originY) + \originY}
        \let\rotY\pgfmathresult
        
        \fill[orange!80] (\rotX,\rotY) circle (0.015);
        
        \coordinate (P\x\y) at (\rotX,\rotY);
      }
    }
    
    \foreach \y in {0,...,\gridHeight} {
      \draw[orange!80, thick] 
        plot coordinates {(P0\y) (P1\y) (P2\y) (P3\y) (P4\y) (P5\y) (P6\y) (P7\y) (P8\y) (P9\y) (P10\y)};
    }
    
    \foreach \x in {0,...,\gridSize} {
      \draw[orange!80, thick] 
        plot coordinates {(P\x0) (P\x1) (P\x2) (P\x3) (P\x4) (P\x5) (P\x6) (P\x7) (P\x8) (P\x9) (P\x10)};
    }
    
    \def\rayStartX{0.1} \def\rayStartY{0.9}
    \def\rayEndX{0.95} \def\rayEndY{0.5}
    \def\nSamples{5} \def\rayColor{purple} \def\rayWidth{1.2pt} \def\sampleRadius{0.02}
    
    \draw[\rayColor, thick, ->, line width=\rayWidth] (\rayStartX,\rayStartY) -- (\rayEndX,\rayEndY);
    
    \foreach \t in {0,1,...,\numexpr\nSamples-1} {
        \pgfmathsetmacro{\alpha}{\t/(\nSamples-1)}
        \pgfmathsetmacro{\rayX}{\rayStartX + \alpha*(\rayEndX-\rayStartX)}
        \pgfmathsetmacro{\rayY}{\rayStartY + \alpha*(\rayEndY-\rayStartY)}
        \fill[\rayColor] (\rayX,\rayY) circle (\sampleRadius);
    }
    
    \end{scope}
\end{scope}

\draw[->, gray, line width=1.2pt] (7.8,1.5) -- (8.5,1.5);

\begin{scope}[xshift=0.5\textwidth]

\node[anchor=south west, inner sep=0] (bunnyC) at (0,0)
  {\includegraphics[width=3cm]{bunny_small.png}};

    \begin{scope}[x={(bunnyC.south east)}, y={(bunnyC.north west)}]
    \draw[orange, line width=0.6pt] (0,0) grid[step=0.1] (1,1);
    
    \def\rayStartX{0.1} \def\rayStartY{0.9}
    \def\rayMidX{0.4,0.6} 
    \def\rayMidY{0.75,0.6}
    \def\rayEndX{0.9} \def\rayEndY{0.25}
    \def\nSamples{5} \def\rayColor{purple} \def\rayWidth{1.0pt} \def\sampleRadius{0.02}
    
    \draw[\rayColor, line width=\rayWidth, ->] 
      (0.1,0.9) -- (0.33,0.8) -- (0.55,0.68) -- (0.75, 0.56) -- (0.95,0.4);
    
    \foreach \x/\y in {0.1/0.9,0.33/0.8,0.55/0.68,0.75/0.56,0.95/0.4} {
      \fill[\rayColor] (\x,\y) circle (\sampleRadius);
    }

    \end{scope}
\end{scope}

\draw[->, gray, line width=1.2pt] (12.1,1.5) -- (12.8,1.5);

\begin{scope}[xshift=0.75\textwidth]

\node[anchor=south west, inner sep=0, opacity=0] (bunnyD) at (0,0)
  {\includegraphics[width=3cm]{bunny_small.png}};

\begin{scope}[x={(bunnyD.south east)}, y={(bunnyD.north west)}]

\tikzset{
  neuron/.style={circle, draw=gray!80, fill=gray!20, minimum size=2mm},
  input/.style={circle, draw=blue!70, fill=blue!20, minimum size=1mm},
  output/.style={circle, draw=green!70, fill=green!20, minimum size=1mm},
  box/.style={draw, rounded corners, fill=white},
  conn/.style={gray!60, thin},
}

\def\xInput{0.00}
\def\xEmbed{0.27}
\def\xHidA{0.50}
\def\xHidB{0.65}
\def\xOut{0.90}


\node[input] (x) at (\xInput, 0.50) {$x$};

\node[box, minimum width=0.18, minimum height=0.15] (gamma)
  at (\xEmbed,0.50) {$\psi(\mathbf{x})$};

\draw[conn] (x) -- (gamma.west);

\foreach \i/\yy in {1/0.65,2/0.50,3/0.35} {
  \node[neuron] (h1\i) at (\xHidA,\yy) {};
}

\foreach \i/\yy in {1/0.65,2/0.50,3/0.35} {
  \node[neuron] (h2\i) at (\xHidB,\yy) {};
}

\node[output] (R) at (\xOut, 0.75) {$a$};
\node[output] (G) at (\xOut, 0.50) {$b$};
\node[output] (B) at (\xOut, 0.25) {$c$};


\foreach \i in {1,2,3} {
  \foreach \j in {1,2,3} {
    \draw[conn] (gamma) -- (h1\i);
  }
}

\foreach \i in {1,2,3} {
  \foreach \j in {1,2,3} {
    \draw[conn] (h1\i) -- (h2\j);
  }
}

\foreach \i in {1,2,3} {
  \draw[conn] (h2\i) -- (R);
  \draw[conn] (h2\i) -- (G);
  \draw[conn] (h2\i) -- (B);
}

\end{scope}
\end{scope}

\end{tikzpicture}
\caption{
Deformation-aware training, from left to right: (a) the rest-space voxel grid (b) the warped deformed-space grid with a straight ray (c) the piecewise linear approximation in rest space (d) index $x$ fed to the Eq.~\ref{eq:encodings} encoder and MLP to predict geometric properties ($a, b, c$). 
}
\label{fig:dnif_upper}
\end{figure*}

\section{Methodology}

To support deformable or animated geometry within the LSNIF framework, a straightforward approach would be to treat each deformed configuration as an independent static object and train a separate neural intersection function per pose. While simple, this strategy is impractical in practice: it requires storing and training multiple models, scales poorly with the number of poses, and fails to exploit the strong geometric coherence shared across deformations. Moreover, even training a single model on multiple deformed shapes directly entangles pose variation with geometry, which significantly degrades accuracy.

To overcome these limitations, we adopt a two-space formulation: We define a \emph{rest space}, which serves as the canonical coordinate system used to train the network, and a \emph{deformed space}, which represents the geometry after a user-specified deformation is applied. The neural network always operates in rest space, while ray casting during dataset generation is performed in deformed space.

We voxelize the geometry in the rest space and apply parameterized deformation to obtain deformed voxels. For each deformation pose, we build a simple BVH over the deformed voxels and trace rays directly in this domain, where rays remain straight lines. Each ray-voxel intersection produces a hit point expressed in deformed-space coordinates. To make this information compatible with the rest-space LSNIF representation, every deformed-space hit point is transformed back into rest space using the inverse deformation. 

Under an exact inverse mapping, a straight ray in deformed space corresponds to a curved path in rest space, and computing the intersection of such curved rays with rest-space geometry would be computationally expensive. In practice, we avoid this cost by applying a local linear approximation. Since LSNIF operates at the voxel level, we assume that within a single voxel, the deformation field is approximately linear. We therefore approximate the curved ray segment inside each voxel by a straight ray in rest space, which allows us to reuse standard ray-voxel and ray-geometry intersection routines. This approximation is efficient and empirically sufficient due to the small spatial extent of individual voxels.
Using this approximation, we intersect the locally linearized rays with the rest-space geometry to obtain ground-truth attributes for training, including hit distance, surface normal, albedo, material, and occlusion. By sampling a wide range of deformation parameters and repeating this process, we generate a dataset that associates deformed-space ray queries with rest-space geometric supervision. 

During training, the transformed hit points in rest space are first encoded using the proposed input encoding described in Eq.~\ref{eq:encodings}. The retrieved grid features are concatenated and passed to a multi-layer perceptron (MLP), which predicts the same set of attributes as the original LSNIF. We employ task-specific loss terms and adopt an uncertainty-weighted multi-task formulation that dynamically adjusts the contribution of each loss, which will be discussed in the following subsections.
This two-space deformation strategy enables a single neural model to remain valid across a wide range of poses. By mapping deformed-space ray queries back to a canonical rest-space representation and leveraging local linearization, the deformation-aware LSNIF retains the compactness and efficiency of the original method while providing stable and accurate intersection prediction for deformable and animated objects without per-pose retraining. The process is illustrated in Figure~\ref{fig:dnif_upper}.

\subsection{Voxel Traversal under Deformation}

Voxel traversal is used to generate ray-voxel intersection samples for training. Unlike uniform grids, deformation induces spatially varying voxel sizes, which makes standard DDA traversal with a fixed step size invalid. Instead, traversal is performed by explicitly determining the voxel face through which a ray exits, which allows correct selection of neighboring voxels even under non-uniform or deformed voxel layouts.

Another challenge is identifying the initial voxel containing the ray origin. Under deformation, voxel indices cannot be computed analytically from the ray origin due to non-uniform voxel extents. Rather than performing an expensive search, we initialize traversal by intersecting the ray with the occupied deformed voxels using a simple BVH. The first intersection point defines the starting voxel for traversal. Although this introduces an additional BVH query, its logarithmic cost is negligible compared to brute-force voxel lookup.

\subsection{Segmented Ray Distance Accumulation}

When a straight ray in deformed space is mapped back to rest space, it becomes piecewise linear across voxels. To compute intersection distances consistently, we accumulate the lengths of these ray segments.

Let $(\mathbf{x}_i^0, \mathbf{x}_i^1)$ denote the entry and exit points of ray segment $i$ within occupied voxel $i$ in deformed space, with segment length $\ell_i = \|\mathbf{x}_i^1 - \mathbf{x}_i^0\|$ and inter-voxel offset $\delta_i = \|\mathbf{x}_i^0 - \mathbf{x}_{i-1}^1\|$. The local hit distance $t_i \in [0, \ell_i]$ is measured from the segment entry point. The total intersection distance is then:
\[
D = \sum_{j=0}^{k-1} (\delta_j + \ell_j) + \delta_k + t_k,
\]
where $k$ indexes the first voxel containing a hit.  

During evaluation, we reconstruct hit positions along the same segmented ray in rest space. This formulation ensures that distance predictions are consistent with spatial localization and simplifies supervision for deformation-aware training.

\subsection{Combining Positional Encoding with the Grid Encoding}

LSNIF represents spatial locations using a multi-resolution hash grid encoding, which provides a compact and efficient alternative to dense voxel grids. While effective for capturing localized geometric detail, we observe that under deformation, relying solely on grid-based features can limit the network's ability to model smooth global variations and high-frequency spatial signals consistently across poses. To address this limitation, we augment the original LSNIF representation with an explicit positional encoding and combine it with the grid encoding in a joint framework.

Given a 3D query point $\mathbf{x} \in \mathbb{R}^{3}$, we apply a sinusoidal positional encoding $\gamma(\mathbf{x})$ defined in
Eq.~\ref{eq:encodings}. The encoding maps each spatial coordinate to a set of sinusoidal functions with exponentially increasing frequencies, where $F$ denotes the number of frequency bands. This formulation injects multi-scale spatial information and enables the network to represent both low-frequency global structure and high-frequency surface detail.

In parallel, we employ the multi-level hash grid encoding. Given a normalized query point $\mathbf{x} \in [0,1]^3$, features are queried from a hierarchy of hash grids with exponentially increasing resolutions. At each level $l$, features are interpolated from neighboring grid vertices and aggregated across levels via concatenation, as formalized in Eq.~\ref{eq:encodings}, where $\phi(\cdot)$ denotes the grid encoder.

We combine positional encoding and hash grid encoding through the composite mapping $\psi(\mathbf{x})$ defined in Eq.~\ref{eq:encodings} \cite{PECS}. Specifically, each query point is first expanded via positional encoding, and the resulting frequency-modulated samples are independently processed by the multi-level hash grid encoder. The encoded features are then flattened and passed to the neural
intersection network.
This joint encoding strategy preserves the locality and memory efficiency of hash grids while retaining the expressive power of explicit sinusoidal features. Empirically, we find that this combination improves robustness under deformation and leads to more stable and accurate prediction of intersection-related
attributes across varying poses.

\begin{equation}
\label{eq:encodings}
\begin{aligned}
\gamma(\mathbf{x}) &=
\left[
\mathbf{x},
\{\sin(2^k \mathbf{x}), \cos(2^k \mathbf{x})\}_{k=0}^{F-1}
\right], \\
\phi(\mathbf{x}) &= \bigoplus_{l=1}^{L} \phi_l(\mathbf{x}),
\qquad
\psi(\mathbf{x}) = \phi\bigl(\gamma(\mathbf{x})\bigr)
\end{aligned}
\end{equation}

\subsection{Uncertainty-Weighted Multi-Task Loss}

We adopt an uncertainty-weighted multi-task objective~\cite{8578879}, where each task has a learnable log-variance $s_i = \log \sigma_i^2$ that dynamically balances its contribution. Regression tasks use $\mathcal{L}_i^{\text{reg}} = 0.5\,\exp(-s_i)\,\mathcal{L}_i + 0.5\,s_i$, classification tasks use $\mathcal{L}_i^{\text{cls}} = \exp(-s_i)\,\mathcal{L}_i + 0.5\,s_i$, and the total loss sums over all tasks.

For distance prediction, deformations produce multi-segment rays, so accumulated distance is unbounded. We replace the sigmoid output with ReLU and regress log-transformed distances using MSE in log space. Surface normals use a cosine similarity loss, albedo is predicted via sigmoid RGB values with $\ell_1$ loss; both as regression tasks. Material is multi-class, occlusion binary; both follow the classification weighting $\mathcal{L}^{\text{cls}}$.

\section{Experimental Results and Discussion}

\subsection{Accuracy Comparison without Deformation}

\begin{figure}[t]
    \centering
    \parbox[t]{0.32\linewidth}{\centering
        \includegraphics[height=3.7cm, keepaspectratio]{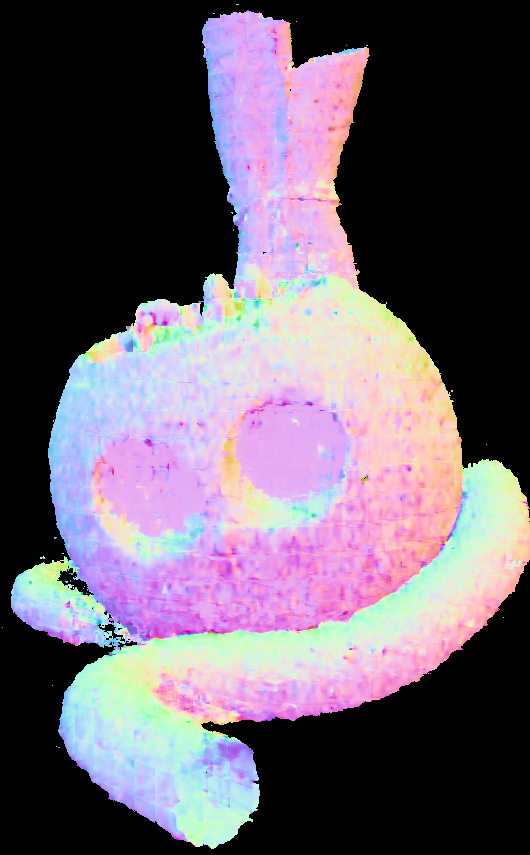}\\
        \small (a) LSNIF}
    \hfill
    \parbox[t]{0.32\linewidth}{\centering
        \includegraphics[height=3.7cm, keepaspectratio]{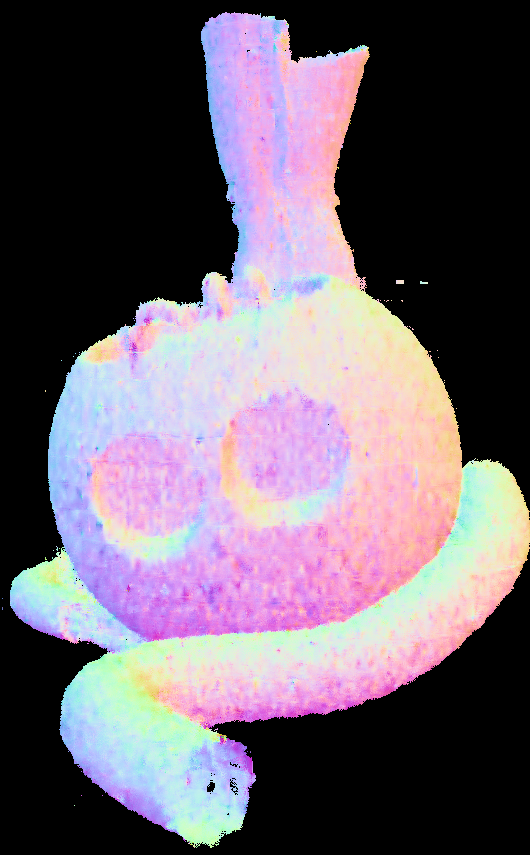}\\
        \small (b) Deformed}
    \hfill
    \parbox[t]{0.32\linewidth}{\centering
        \includegraphics[height=3.7cm, keepaspectratio]{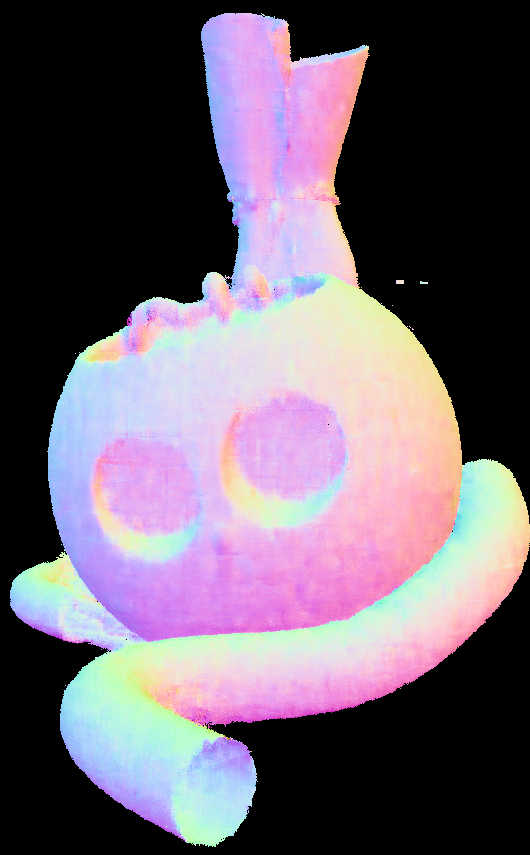}\\
        \small (c) Proposed}

    \caption{Single-pose comparison: (a) original LSNIF, (b) deformation-aware without our modifications, (c) deformation-aware with all proposed methods.}
    \label{fig:lsnif_vs_deformed_images}
\end{figure}

To isolate the effect of deformation-aware training, we evaluate both our method and the original LSNIF on static, single-pose geometry. No deformation is applied, so differences arise solely from training strategy. As expected, the original LSNIF slightly outperforms in this setting, since training on a single pose allows overfitting to pose-specific geometry and appearance, yielding lower distance and normal errors (Table~\ref{tab:lsnif_vs_deformed}). Our deformation-aware model, trained on a pose distribution, is regularized and cannot fully specialize to a single configuration. Nevertheless, the proposed modifications reduce this gap, and in some metrics our model even surpasses LSNIF. Figure~\ref{fig:lsnif_vs_deformed_images} shows a visual comparison.

\begin{table}[t]
\centering
\caption{Prediction error on a single static pose. 
\textbf{LSNIF} denotes the original method trained on a single pose.
\textbf{Deformed} is a deformation-aware baseline without the proposed training and encoding modifications.
\textbf{Proposed} applies all proposed methods.}
\label{tab:lsnif_vs_deformed}
\begin{tabular}{lcc}
\hline
Method & Distance (mean-absolute-error) & Normal ($^\circ$) \\
\hline
LSNIF    & 0.105 & 19.1 \\
Deformed & 0.135 & 24.7 \\
Proposed & \textbf{0.039} & \textbf{11.8} \\
\hline
\end{tabular}
\end{table}



\subsection{Comparison of Deformation Results}

\begin{figure}[t]
    \centering
    \parbox[t]{0.24\linewidth}{\centering
        \includegraphics[height=2.7cm, keepaspectratio]{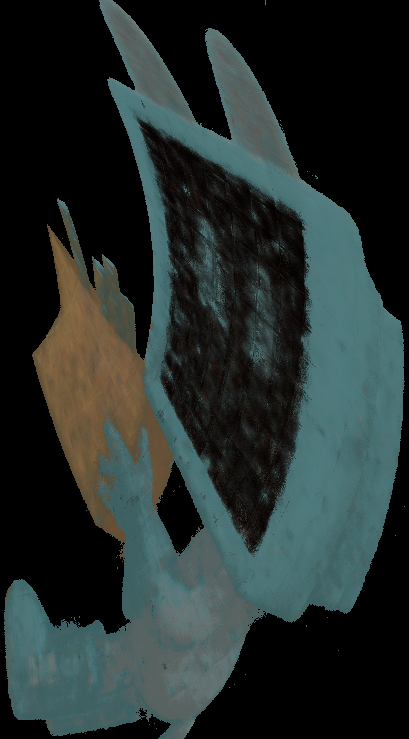}\\
        \small (a1) Deformed}
    \hfill
    \parbox[t]{0.24\linewidth}{\centering
        \includegraphics[height=2.7cm, keepaspectratio]{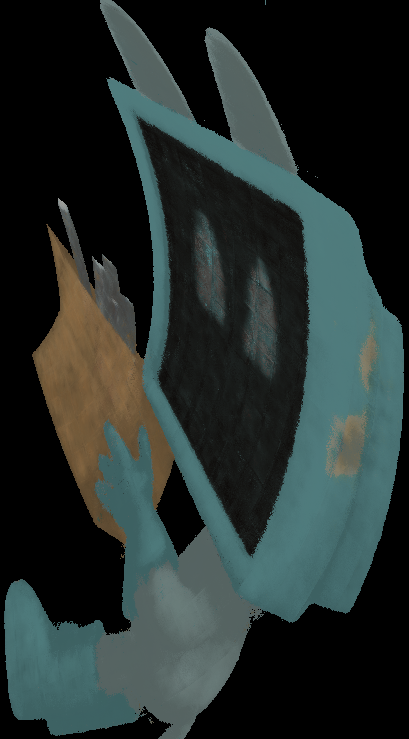}\\
        \small (b1) Proposed}
    \hfill
    \parbox[t]{0.24\linewidth}{\centering
        \includegraphics[height=2.7cm, keepaspectratio]{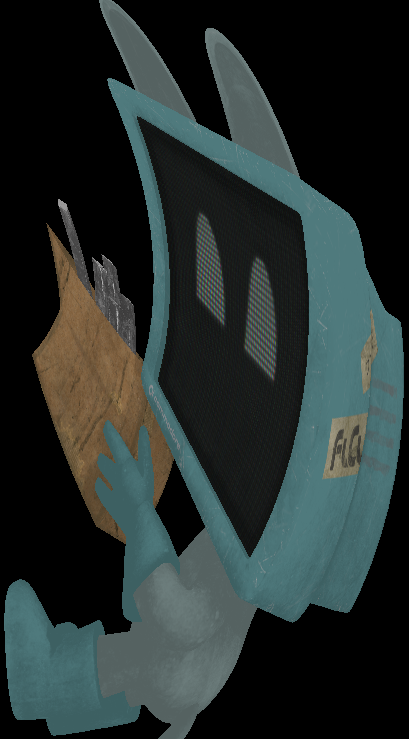}\\
        \small (c1) Reference}
    \hfill
    \parbox[t]{0.24\linewidth}{\centering
        \includegraphics[height=2.7cm, keepaspectratio]{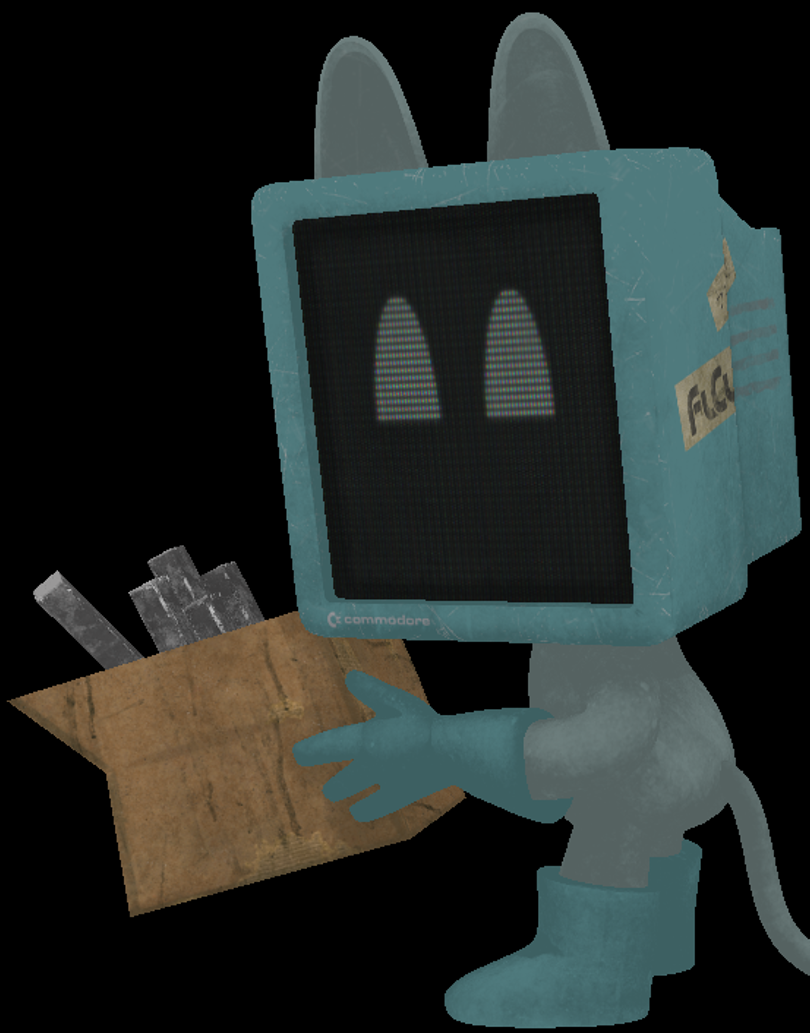}\\
        \small (d1) Original}

    \vspace{4pt}

    \parbox[t]{0.24\linewidth}{\centering
        \includegraphics[height=2.7cm, keepaspectratio]{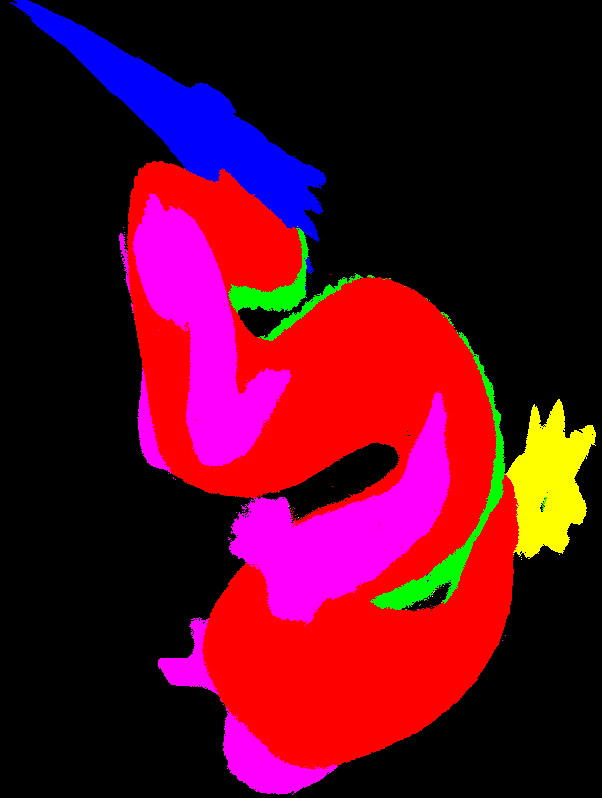}\\
        \small (a2) Deformed}
    \hfill
    \parbox[t]{0.24\linewidth}{\centering
        \includegraphics[height=2.7cm, keepaspectratio]{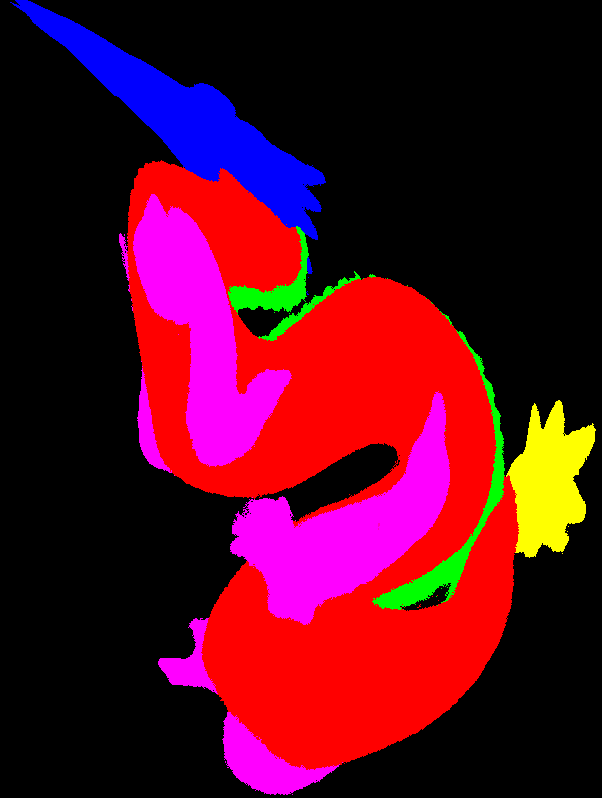}\\
        \small (b2) Proposed}
    \hfill
    \parbox[t]{0.24\linewidth}{\centering
        \includegraphics[height=2.7cm, keepaspectratio]{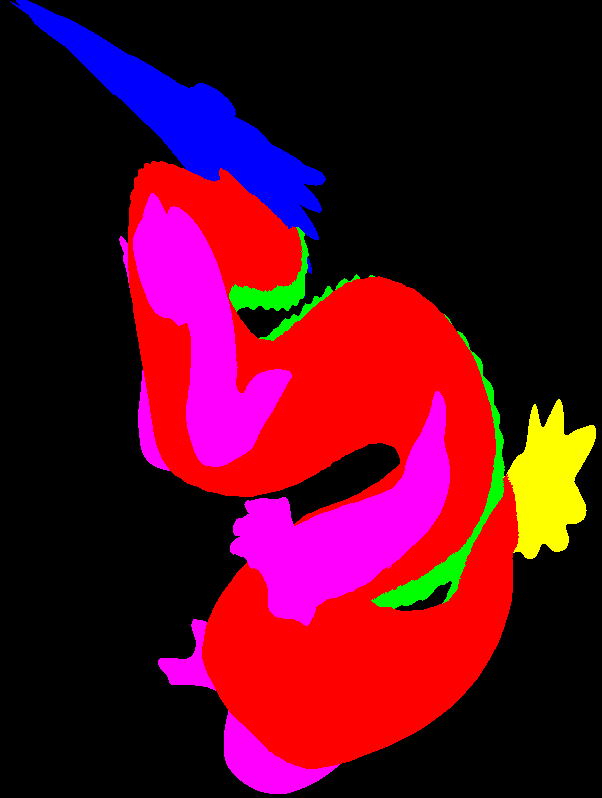}\\
        \small (c2) Reference}
    \hfill
    \parbox[t]{0.24\linewidth}{\centering
        \includegraphics[height=2.7cm, keepaspectratio]{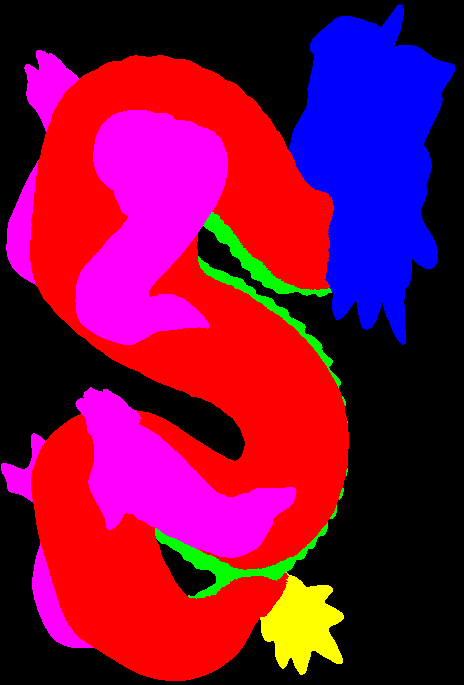}\\
        \small (d2) Original}

    \caption{Comparison of deformation-aware predictions across two scenes. 
    Top row: albedo prediction on the cat object. 
    Bottom row: multi-material prediction on the dragon object.}
    \label{fig:combined_deformed_results}
\end{figure}

We compare a deformation-aware LSNIF baseline without our modifications to the full model incorporating all proposed components. Both are trained on the same deformed poses and evaluated on identical ray queries with voxel resolution $32^3$, a two-level hash grid (base 64, feature dim 3), and $F=3$ positional encoding bands. By employing a simple BVH over deformed voxels and trace straight rays directly in this domain, where rays remain straight; this adds only 2--3\% computational overhead. As shown in Figure~\ref{fig:combined_deformed_results}, the baseline suffers degraded intersection accuracy and appearance consistency under deformation, with blurred surfaces and unstable attributes. The full model produces sharper, consistent geometry, demonstrating the importance of our deformation-aware extensions for robust neural intersection prediction.

\section{Conclusion, Limitation, and Future Work}

We presented a deformation-aware extension of LSNIF that enables efficient neural intersection prediction for parameterized deformable geometry. By mapping ray queries from deformed space to a canonical rest space, a single neural model generalizes across multiple poses without per-pose retraining. Curved rays induced by deformation are approximated using voxel-wise linear segments, preserving accuracy and efficiency.

To improve stability under pose variability, we introduced log-scaled distance regression, uncertainty-weighted multi-task learning, and a hybrid positional-grid encoding. Experiments show that neural intersection functions can be extended beyond static geometry while retaining LSNIF's compactness.

While our approach handles articulated and simple morphing deformations, highly non-linear, topologically changing, or physics-driven deformations may require more expressive representations or adaptive voxelization. Future work includes supporting these general deformations, learning deformation fields jointly with intersection prediction, and integrating the method into full rendering pipelines.

\bibliographystyle{eg-alpha-doi}
\bibliography{deform_ref.bib}

\end{document}